# Coalition Game-based Approach for Improving the QoE of DASH-based Streaming in Multi-servers Scheme

Oussama El Marai*, Miloud Bagaa*, and Tarik Taleb*§
oussama.elmarai@aalto.fi, miloud.bagaa@aalto.fi, tarik.taleb@aalto.fi
*Aalto University, Espoo, Finland.
§University of Oulu, 90570 Oulu, Finland.

*Abstract*—Dynamic Adaptive Streaming over HTTP (DASH) is becoming the de facto method for effective video traffic delivery at large scale. Its primer success factor returns to the full autonomy given to the streaming clients making them smarter and enabling decentralized logic of video quality decision at granular video chunks following a pull-based paradigm. However, the pure autonomy of the clients inherently results in an overall selfish environment where each client independently strives to improve its Quality of Experience (QoE). Consequently, the clients will hurt each other, including themselves, due to their limited scope of perception. This shortcoming could be addressed by employing a mechanism that has a global view, hence could efficiently manage the available resources. In this paper, we propose a game theoretical-based approach to address the issue of the client's selfishness in multi-server setup, without affecting its autonomy. Particularly, we employ the coalitional game framework to affect the clients to the best server, ultimately to maximize the overall average quality of the clients while preventing re-buffering. We validate our solution through extensive experiments and showcase the effectiveness of the proposed solution.

## I. INTRODUCTION

During the current decade, there was a tremendous development in hardware devices, notably mobile devices, in terms of processing and display capabilities. Besides, the spectacular improvement in network bandwidth brought by 4G networks has dramatically increased and stimulate the user's engagement for watching videos over best-effort networks. The coming 5G network is promising further enhancements in terms of speed and lowering latency in order to accommodate the ever-growing traffic demands. All these factors together, alongside the phenomenal success and proliferation of over-the-top (OTT) services, have fairly lead to the dominance of video traffic over the network.

According to many studies [1], video traffic is impressively growing at a high rate and is currently dominating the internet traffic in all networks, notably mobile and wireless networks. This would exert a considerable burden on the underlying infrastructure, even though the continuous advances and improvements, since the users' consumption increases accordingly. As a result, guaranteeing an end-to-end QoS is almost not feasible especially in best-effort networks. This calls for employing intelligent and flexible solutions that are able to self-adapt to the network dynamics.

To cope with the network limitation and provide users with the highest possible QoE, Dynamic Adaptive Streaming over HTTP (DASH) has appeared and became the protocol of choice for delivering both Video on Demand (VoD) and live streams. Presently, DASH technique is gaining surging popularity and being widely adopted by many giant OTT services (e.g. YouTube and Netflix). The success of DASH stems from the plenty advantages it exhibits, notably its ability to dynamically adapt to network changes. This can be achieved by encoding the raw video into multiple video qualities with different encoding rates called representations. Each version is then divided into non-overlapping segments of short duration, typically between 2 and 10 seconds. At the client-side, the video segments are requested back to back with separate HTTP GET requests. After each segment's reception, the client evaluates its reward function and decides the representation of the next segment accordingly. The different downloaded segments are held into a buffer for further playout since the TCP throughput is inherently sawtooth and unpredictable. At each segment boundary, the client can shift up or down, within the available video quality set, depending on the adaptation policy.

Despite the aforementioned fascinating features, many studies revealed that DASH is actually suffering a number of shortcomings. For instance, the autonomy given to the clients can indeed backfire (double-edged) and unleashes clients' selfishness. This selfishness would hurt everyone, including themselves, and might be deleterious for the overall system performances. For instance, the instability issue is one of the side effects of selfishness since each client tries to maximize its perceived video quality without taking into consideration the other clients. This behavior would result in throughput misconception.

Game Theory (GT) is a branch of mathematics that enables the modeling and analysis of the interactions between several decision-makers (called players) who can have conflicting or common objectives [2]. This branch offers a number of frameworks and models that motivate the formulation of DASH-related issues as a game between the different system players. These frameworks seem suitable to be employed in DASH context where we have multiple selfish clients taking decisions (i.e. the bitrate of the next chunk) and competing for network resources. This competitiveness and interaction with the network play a big role in the resulting individual and overall performance of the system.

In this paper, we propose a new GT-based solution that improves both the per-client and overall average QoE in a multi-servers scheme, where the fetched content is redundant in multiple servers (e.g. CDN servers). In the proposed solution, the clients form coalitions and move from one coalition to another based on the coalitions' payoff. The payoff is a function that measures the servers' load and allows the clients to migrate, based on their selected video quality and the leftover bandwidth at the server, to the best possible server that provides the clients with more bandwidth room to improve their video quality. Experimental results, show the effectiveness of the proposed solution compared to the random attribution of the clients to the different servers.

The rest of this paper is structured as follows: Section II presents the related work. In section III, we formulate the problem and provide a detailed description of the GT model as well as the proposed solution. In Section IV we describe the experimental setup and parameters as well as the relevant results. Finally, Section V summarizes our findings and highlights future research directions.

## II. RELATED WORK

Since its appearance in 2011, DASH technology attracted many researchers from academia as well as industry. Consequently, plenty of work has been proposed to ultimately provide the user with the best possible QoE. In this section, we provide the crux of the most relevant proposed work, shedding the light on the controlled-based approaches, and we refer the reader to [3], [4] for a thorough review.

In [5], the authors proposed a game-theoretical approach to achieve a high and stable user QoE, for VoD streams, by employing a bargaining game and consensus mechanism in the design and formulation of the decision problem. Depending on the number of players, the proposed approach employs either a collaborative or non-collaborative approach. With the existence of concurrent players in a shared environment, the proposed system employs the collaborative approach, whereas the other mode is used when only one player is connected. The proposed approach has been implemented in real-world and its performance is compared against the existing state of the art solutions. The authors reported that the proposed approach improves the average QoE by 38.5%, the stability by 62%, prevents rebuffering events, and reduces the startup delay.

Yuan et al. formulate in [6] the problem of concurrent bandwidth access in DASH context using a purely non-cooperative game theoretical approach. The proposed approach aims to address the users' selfishness behavior, causing conflicts among the competing players, and resulting in sub-optimal perceived video quality and unfair bandwidth allocation. To address the conflict, the authors propose a novel rate adaptation algorithm that improves the user's QoE, while ensuring fairness among the players. The algorithm selects the video chunks based on the local information available at the player side and the global payoff received from the server. The existence of a Nash Equilibrium of the game has been proven too using a distributed iterative algorithm with stability analysis. Additionally, the authors proposed a novel QoE model that incorporates various parameters including the actual and reference buffer lengths, and the perceived video quality. In spite of the achieved results in terms of video quality and fairness guarantee, the proposed algorithm suffers from a major issue consisting of opening additional HTTP sessions between the server and the clients, which may question its efficiency at large scale.

The authors in [7] proposed a Fully Distributed Collaborative HAS (FDCHAS) scheme for VoD services, based on game theory framework with consensus mechanism, to address the scalability of the adaptation algorithms by avoiding explicit communication between HAS components. The proposed solution entails a two-stage game, residing at client and network sides respectively. The effectiveness of the proposed scheme has been validated through trace-driven and real-world experiments. Besides, a comparison to the previous state of the art solution shows the improvement of the user QoE, stability, and fairness.

Probe AND Adapt (PANDA) [8] proposed by Li et al. is a rate-based client-side algorithm that aims at eliminating the inherent ON-OFF effect of the segments downloading process, and provides a stable viewing experience. PANDA is composed of four main steps namely bandwidth estimation, smoothing the estimated bandwidth using the harmonic mean, select the video bitrate of the next segment, and finally schedule the next download. The proposed algorithm shows good results in terms of bitrate stability. However, it seems to be very conservative to the point that it does not switch to the next level even though the available bandwidth allows it.

SDN assisted Adaptive BitRate (SABR) architecture has been proposed in [9] by Bhat et al. to feed DASH clients via REST APIs with measurements (e.g. cache occupancy, available bandwidth) on the network conditions. The proposed scheme leverages the SDN paradigm and uses the received information from the network to increase the clients' awareness and obtain more accurate information, ultimately for the improvement of the user's QoE. In [10], the authors proposed a multi-objective model for segment-based path selection in SDN-enabled networks for providing high QoE based on the available bandwidth, the segment bit-rate, and the path length. The obtained results of the proposed approach show better performances compared to the traditional best-effort routing.

In this paper, we aim to alleviate the impact of the clients' selfishness in a multi-server setup, without affecting the clients' autonomy, using game theory framework. Differently from previous work, we have found relevant the use of a coalitional game in such a setup where each server represents a coalition of clients. In such a scheme, and thanks to the chunked version of the video which allows individual requests for each video chunk, the clients are not only able to change the video quality at each chunk boundary, but also have the freedom of dynamically requesting the chunks from different sources, mainly for improving their QoE.

## III. COALITIONAL GAME FOR IMPROVING QOE IN MULTI-SERVERS

### A. Main Idea and Problem Formulation

We model the underlying network as a bipartite graph $G(U, S, E, W)$. The graph $G$ mainly interconnects the users (i.e., DASH players, denoted by $U$, with the set of servers $S$. While $E$ denotes the edges (i.e., Network edges) that connect $U$ with $S$. $W$ denotes the end-to-end bandwidth between the users $U$ and the servers $S$. For the sake of simplicity, $E$ represents the end-to-end connections between $U$ and $S$ by considering an overlay network and by making an abstraction on the intermediate routers and switches (backbone). However, the suggested model is orthogonal and can be easily adapted and extended to consider the intermediate network components. Based on the observation that DASH streaming approach is a completely distributed system, and the behavior of the client is mainly affected by the network dynamics, therefore, the clients constantly measure their bandwidth. Let $B_s$ denote the bandwidth of the server $s \in S$. Let $\lambda_u$ denote the requested video bitrate of the client $u \in U$.

We have leveraged the coalitional game to improve the QoE of DASH-based clients in a distributed multi-server environment. The objective of the game is to enhance the perceived QoE for each client by optimally assigning the bitrates and servers. The coalitional games are classified into three classes [11]: i) Canonical (coalitional) games; ii) Coalitional graph games; iii) Coalitional formation games. In the former, the grand coalition is optimal, and the aim is to stabilize the grand coalition by providing an appropriate payoff vector. In the second class, the players interact among themselves using a dedicated graph. This class aims to stabilize the grand coalition or various coalitions while considering the communication graph. In other words, two players can interact if and only if they are neighbors in the communication graph. Last but not least, coalitional formation games class, which is adopted in this paper, aims to form different coalitions, such that the profits of the players are increased.

Formally, we define the game $G(P, C, F)$ as follows: $P$ is the set of players in the system, which are the DASH clients. Formally, $P = U$. Meanwhile, $C$ is the set of coalitions in the game. In the game, the clients that use the same server form one coalition. Thus, the number of coalitions is fixed and exactly equals the number of servers in the system, and hence $C = S$. Let $N$ denote the cardinality of the servers in the system. Therefore, $C = \{C_1, C_2, \cdots, C_N\}$, such that $C_i$ denotes the coalition of clients that use the same server $i$. Note that every two coalitions have a different set of players. $C_i, C_j \in C : i \neq j \Rightarrow C_i \cap C_j = \emptyset$. Let $\Sigma$ denote the grand coalition that consists of all the players in the game. Formally, $\bigcup_{i \in C} C_i = \Sigma$. Meanwhile, $F$ denotes the characteristic function that defines the payoffs of the players of the coaltions $C$. Let $F(C_i)$ denote the characteristic function of a coalition $C_i \in C$.

### B. Solution Description

In this subsection, we describe the proposed solution that aims to leverage coalitional games for efficiently enhancing the QoE of DASH-based clients in a multi-server system. In the balance of this section, we will first present an overview of the used game. Then, we define rules to ensure smooth clients transfer between different coalitions. Finally, we describe the proposed solution.

In the system, the requested quality per each client is fixed from one side, and from another side each player wants to increase his benefit by selecting the coalition (i.e., server) that offers more chance to enhance the perceived quality in the future. We define the payoff of the coalition as the difference between the requested quality and the remaining bandwidth at each server. Formally, the server that will offer more quality in the future is the one that has a higher chasm between its network bandwidth and the requested bitrates of its clients. The more the chasm between the server's bandwidth and the requested qualities of the clients, the more likely to offer better qualities for the clients connected to that server. We define the characteristic function of the coalitional game as follows

$$\forall C_i \in C : F(C_i) = B_{C_i} - \sum_{p \in C_i} \lambda_p \quad (1)$$

---

**Algorithm 1:** Game Theory Algorithm

1 initialization;
2 **while** *True* **do**
3     $migrated \leftarrow False$;
4     **foreach** $C_i \in C$ **do**
5        **foreach** $p \in C_i$ **do**
6           **foreach** $C_j \in C \setminus C_i$ **do**
7              **if** CHECK_MIGRATION($p, C_i, C_j$) **then**
8                 $migrated \leftarrow True$;
9                 *break*;
10              **end**
11           **end**
12           **if** *migrated=True* **then**
13              *break*;
14           **end**
15        **end**
16     **if** *migrated=True* **then**
17        *break*;
18     **end**
19     **end**
20     **if** *migrated=False* **then**
21        *break*;
22     **end**
23 **end**

---

In our coalitional formation game, we have considered the equal-sharing method to share the benefit among the players within the same coalition. However, any other method (i.e., shapely value or nucleolus) can be used in the game with slight modification. In the coalitional game, the players keep moving from a coalition to another to enhance their benefit. Our game is individual rationality, which means that no player would move to another coalition, whereby he will receive a payoff less than the one that he is currently receiving. Thus, the player should move from a coalition to another if and only if he will receive a better payoff. In our game, we define the user transfer operation that enables the player to move from one coalition to another to increase his payoff. Let $C_i$ and $C_j$ denote two coalitions in our framework, thus when a player $p \in C_i$ moves from the collation $C_i$ to the collation $C_j$, then we will end up with two new collations $\overline{C_i^p}$ and $\hat{C}_j^p$. For the sake of example, let consider that the collection consists only of two coalitions, thus $\Sigma = C_i \cup C_j$ and $C_i \cap C_j = \emptyset$. Using the transfer operation, we ensure that the movement from a collection $\{C_i, C_j\}$ to another collection $\{\overline{C}_i^p, \hat{C}_j^p\}$ ensures the individual and coalitional rationality. The transfer is effective if and only if the perceived payoffs of the resulting coalitions is better than the original ones.

**Definition 1.** *Transfer rule* A player $p \in C_i$ would be transferred to another coalition $C_j$, resulting two new coalitions $\overline{C}_i^p, \hat{C}_j^p$ if and only if:

$$\{C_i, C_j\} \, d^p \, \{\overline{C}_i^p, \hat{C}_j^p\} \iff F(\overline{C}_i^p) > F(C_i) \land F(\hat{C}_j^p) > F(C_j). \quad (2)$$

In (2), we denote by $d^p$ the transfer superiority of the new collection $\{\overline{C}_i^p, \hat{C}_j^p\}$ than the collection $\{C_i, C_j\}$. From (2), the player $p$ should be transferred iff the following statements hold: The first statement, i.e., $F(\overline{C}_i^p) > F(C_i)$, obviously holds due to the superiority of the perceived gain of $\overline{C}_i^p$ than $C_i$. In fact, the remove of the player $p$ from $C_i$, resulting a new coalition $\overline{C}_i^p$, will alleviate the overhead and then provide a higher payoff. Meanwhile, the second statement, $F(\hat{C}_j^p) > F(C_j)$, means that the transfer should happen if and only if the gain that we perceive in the new coalition (i.e., $\hat{C}_j^p$) is higher than the perceived from the original coalition (i.e., $C_j$). This strategy will enhance the likelihood to perceive higher quality for the clients in the future. From the above, we conclude that the transfer should happen only if the gain is higher than the loss.

**Definition 2.** *A collection of coalitions is stable if and only if no further transfer can be applied. Formally, the following statement should hold:*

$$\forall C_i, C_j \in C, C_i \cap C_j = \emptyset, \forall p \in C_i \implies \nexists \overline{C}_i^p, \hat{C}_j^p \in C,$$
$$\overline{C}_i^p \cap \hat{C}_j^p = \emptyset,$$
$$\overline{C}_i^p \cup \hat{C}_j^p = C_i \cup C_j,$$
$$\{C_i, C_j\} \, d^p \, \{\overline{C}_i^p, \hat{C}_j^p\} \quad (3)$$

According to the statement (3), a collection of coalitions $C$ is stable if and only if: For all two coalitions $C_i, C_j \in C$, it should not be exist any player that moves from $C_i$ to another coalition $C_j$, forming new two coalitions, $\overline{C}_i^p$ and $\hat{C}_j^p$, where the players perceive better payoffs. The part, $\nexists \overline{C}_i^p, \hat{C}_j^p \in C$, in statement (3) ensures that there should be no combination $(\overline{C}_i^p, \hat{C}_j^p)$ where the player $p$ moves from previous combination $(C_i, C_j)$ that have better payoff. The latter is ensured by the last part of the inequality, which is $\{C_i, C_j\} \, d^p \, \{\overline{C}_i^p, \hat{C}_j^p\}$. Meanwhile, the remaining two parts, $\overline{C}_i^p \cap \hat{C}_j^p = \emptyset$ and $\overline{C}_i^p \cup \hat{C}_j^p = C_i \cup C_j$ are obvious and ensured by the first part of the inequality.

**Theorem 1.** *The proposed Algorithm terminates and produces an $ID_p$-stable collection and provides optimal configuration.*

*Proof.* Our Algorithm is executed in steps, such that at each step the Algorithm moves from a collection to another. Let $\Gamma^i$ denote the collection at the step $i$ in the Algorithm. At each step, $\Gamma^i = \{C_1^i, C_2^i \cdots\}$. Then, the Algorithm keeps moving from a collection to another until the convergence and providing the optimal configuration.

$$\Gamma^0 \to \Gamma^1 \to \Gamma^2 \to \Gamma^3 \ldots \quad (4)$$

, where $\Gamma^0$ denotes the initial collection that is randomly generated, and $\to$ reflect the transformation process that happens due to players' transfer from a collection to another.

Our Algorithm keeps moving from a collection ($\Gamma^i$) to another $\Gamma^{i+1}$ using the inequality (2). The transformation processes keep applying from a collection to another until the statement (3) holds. Based (2), we will move from a collection ($\Gamma^i$) to the next collection ($\Gamma^{i+1}$) if and only if: At least the payoff of one coalition $\hat{C}_j^p$ is better than the coalition $C_i$ in the previous collection by transferring the player $p$. Thus, there is no possibility to move from a collection to another whose coalition has a lower payoff. Based on the observation that there is no cycle in the coalition generation, and the number of players and servers is limited, our Algorithm finishes and provides an $ID_p$-stable collection that provides an optimal configuration, which respects the statement mentioned in (3). □

---

**Algorithm 2:** Check Migration Algorithm

1 $F(C_i) = B_{C_i} - \sum_{p \in C_i} \lambda_p;$
2 $F(C_j) = B_{C_j} - \lambda_p - \sum_{p \in C_j} \lambda_p;$
3 **if** $F(C_j) - F(C_i) \geq \xi$ **then**
4     $F(C_i) \leftarrow F(C_i) \setminus \{p\};$
5     $F(C_j) \leftarrow F(C_j) \cup \{p\};$
6     return True;
7 **end**
8 retrun False;

---

*C. Algorithm Description*

In this subsection, we detail the proposed algorithm that performs the different combinations to reach the $ID_p$-stable collection. The initialization phase (line 1) launches a defined number of clients and affects them randomly to the available DASH servers, which forms the initial collection $\Gamma^0$. The outer loop is used to ensure iterating through all existing coalitions (i.e., servers) and their corresponding players (i.e., clients). A boolean variable *migrated* is initialized to *False*, which helps to indicate whether migration has been performed successfully or not. It allows restarting the iterations over the existing clients and servers each time migration is performed. Then, we loop through the existing servers (line 4), and for each client belonging to the actual server (line 5), we check a possible transfer of that client to another server. To do that, we need to iterate over the available servers except for the server that the current client belongs to (line 6). After that, the function $check\_migration(p, C_i, C_j)$ is called to effectively perform the migration of the current player $p$ from $C_i$ to $C_j$ if the inequality (2) holds. In this case, it returns *True* and consequently, the Algorithm exits the current loop and all other loops except the outer one. Otherwise, we just continue the iterations over the remaining players and coalitions. The pseudo-code is provided in Algorithm 1.

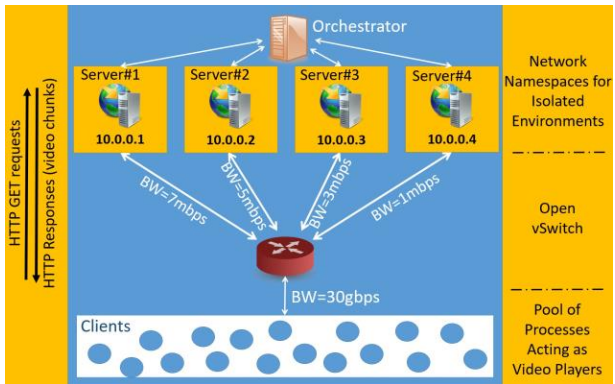

Fig. 1: Experimental setup.

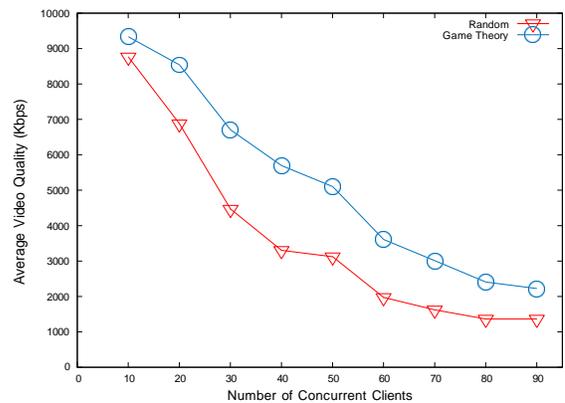

Fig. 2: Average video quality when varying the number of simultaneous clients.

## IV. PERFORMANCE EVALUATION

### A. Experimental Setup

We have conducted extensive experiments to evaluate the performance of the proposed game theory-based solution against a baseline solution. The latter consists of a random affectation of the launched clients to one of the existing servers. In the present experiments, we have created four separate virtualized servers using the Network Namespaces feature provided in Linux operating systems. These servers are interconnected with each other, as well as to the host machine, via an Open vSwitch (OVS). The video players are launched as independent processes at the host machine and run the DASH-based client algorithm proposed in [12]. The four servers have different bandwidth capacities, by throttling their corresponding links, to demonstrate how the clients' distribution is relatively depending on the servers' capacity, thanks to the payoff function. The servers' capacities are 7Mbps, 5Mbps, 3Mbps, and 1Mbps respectively. The virtual link between the switch and the host machine has a very high capacity (3Gbps) to avoid creating a bottleneck at the client-side. The experimental setup is depicted in Fig. 1.

Several experiments have been conducted by varying the number of simultaneous running clients from 10 to 90 clients, with a step of 10. These clients request a recorded video encoded at different bitrates: 1360Kbps, 3265Kbps, 6117Kbps, 9330Kbps, corresponding to four qualities 360p, 480p, 720p, 1080p , respectively. Basically, the clients perform HTTP GET requests and the servers deliver the requested video chunks with HTTP responses. In our experiments, the orchestrator saves the information regarding the allocation of clients to their corresponding web servers in Redis cache system. Similarly, the clients retrieve the name of the server from which they would request the next chunk from Redis cache memory.

Initially, the clients are randomly attributed to the available servers, where each client is affected to one and only one server at a time, and we measure the average video quality during the whole initial phase. Then, we execute the game theory algorithm and perform the required transfers of the clients from one coalition to another until we reach the $ID_p$-stable collection. It should be noted here that during the ramp-up phase (i.e. when we launch the clients), we do not take into consideration the selected video bitrates to avoid any bias to the results of the game theory phase over the random phase. The reason is that during the startup phase the client's algorithm starts from the lowest video quality and gradually increases the bitrate if the available bandwidth allows it. Additionally, we make sure that the average video quality is calculated at greedy state over the same number of values at both phases. In this paper, we are considering the measurement of the following metrics: *i)* the average video quality of all launched clients; *ii)* the load of each server in terms of attributed clients, and *iii)* the number of migration operations performed at each execution. The game theory algorithm runs on an orchestration server that has a global view on the load of the streaming servers.

### B. Experimental Results

*1) Average Video Quality:* The results of the average perceived video quality are illustrated in Fig. 2. Each data point in this figure shows the average video quality of 10 different executions at different numbers of competing clients. From this figure, we vividly see that the video quality is inversely proportional to the number of concurrent clients. Also, the average video quality after executing the game theory algorithm is notably higher compared to the initial client's attribution.

*2) Clients distribution over servers:* In Fig. 3, we plot the number of affected clients per server at both random and game-theoretical phases. Generally, the random affectation of the clients results in unfair attribution to the servers regardless of their bandwidth capacity. For instance, we see that Server4 having the lowest capacity has more clients than Server1 with the highest bandwidth capacity. After the execution of the proposed algorithm, we notice that the redistribution of the clients is proportional to the servers' bandwidth capacities, which results in fair load distribution and higher perceived video quality as discussed previously.

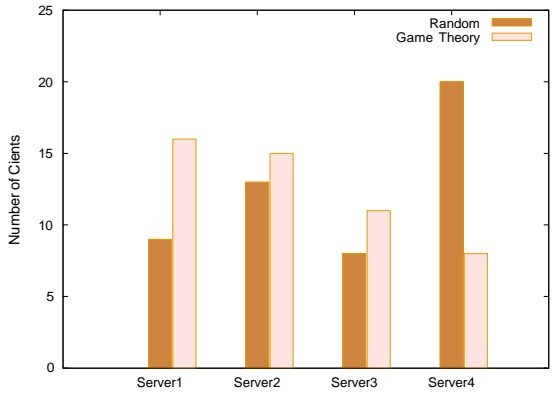

Fig. 3: Server's load in terms of affected clients.

*3) Migrations per execution:* In Fig. 4, we showcase the average number of transfers performed to reach the $ID_p$-stable collection at each different number of concurrent clients. In this figure, each bar represents the average value of ten executions at the given parallel clients. Obviously, we conclude from this figure that the number of transfers required to reach the $ID_p$-stable collection is proportional to the number of active clients in the system.

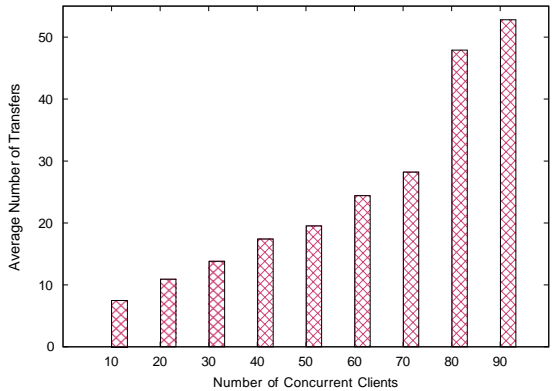

Fig. 4: Average number of transfers.

## V. CONCLUSION

In this paper, we have proposed a coalitional-based game theory approach to address the problem of sub-optimal perceived video quality of DASH-based clients in a multi-servers scheme. This issue is a result of the selfish behaviour of the clients due to their full autonomy in bitrate decision making. The proposed solution consists of forming clients' coalitions, based on the allocated server, and keep migrating the clients from one coalition (i.e. server) to another based on the residue bandwidth at source and destination coalitions. The payoff function consists of the difference between the residue bandwidth between the destination and source servers respectively. The migration is performed only if the difference is greater than a predefined threshold. The migration algorithm runs on a controller that has a global view on the servers' load. Experiment results demonstrate the effectiveness of the proposed solution in improving the average perceived quality while preventing clients from rebuffering.

Although the proposed solution notably improves the average perceived video quality of DASH-based clients in multi-servers setup while offering stall-free playback, the proposed scheme does not take into consideration other metrics that might have detrimental consequences on the user's QoE, such as stability and fairness. Our future work consists of amending the current model by incorporating such lofty goals to the payoff formula and study the overall performances when simultaneously considering multiple design goals.

ACKNOWLEDGEMENT

This work was partially supported by the European Union's Horizon 2020 research and innovation programme under the ACCORDION project with grant agreement No. 871793, and by the Academy of Finland's Flagship programme 6Genesis under grant agreement No. 318927. It was also supported in part by the Academy of Finland under CSN project with grant No. 311654.